\documentstyle[twoside,fleqn,espcrc2]{article}

\newcommand{\AmS}{{\protect\the\textfont2
  A\kern-.1667em\lower.5ex\hbox{M}\kern-.125emS}}

\title{Large Mixing and $CP$ Violation in Neutrino Oscillations}

\author{Zhi-zhong Xing
	\address{Sektion Physik, Universit$\rm\ddot{a}$t M$\rm\ddot{u}$nchen,
Theresienstrasse 37A, 80333 M$\rm\ddot{u}$nchen, Germany}
	\thanks{Talk given at the Sixth Topical Seminar on 
Neutrino and Astroparticle Physics, San Miniato, Italy, May 1999}}
       
\begin{document}

\begin{abstract}
I introduce a simple phenomenological model of lepton flavor mixing
and $CP$ violation based on the flavor democracy of charged 
leptons and the mass degeneracy of neutrinos. The nearly bi-maximal
mixing pattern, which can interpret current data on atmospheric
and solar neutrino oscillations, emerges naturally from this model.
The rephasing-invariant strength of $CP$ or $T$ violation 
amounts to about one percent and could be measured in the 
long-baseline neutrino experiments. 
The similarity and
difference between lepton and quark flavor mixing phenomena 
are also discussed.
\end{abstract}

\maketitle

{\Large\bf 1} ~
In the standard electroweak model neutrinos are assumed to be the 
massless Weyl particles. This assumption, which has no conflict with 
all direct-mass-search experiments \cite{PDG}, is not guaranteed 
by any fundamental principle of particle physics.
Indeed most extensions of
the standard model, such as the grand unified theories of quarks
and leptons, allow the existence of massive neutrinos. If the masses of
three active neutrinos ($\nu_e$, $\nu_\mu$ and $\nu_\tau$) are 
nonvanishing, why are they so small in comparison with the masses
of charged leptons or quarks? For the time being this question remains
open, although a lot of theoretical speculations towards a definite
answer have been made. 
The smallness of neutrino masses 
is perhaps attributed to the fact that neutrinos are electrically neutral
fermions, or more exactly, to the Majorana feature of neutrino fields.

Recent observation of the
atmospheric and solar neutrino anomalies, particularly
that in the Super-Kamiokande experiment, has provided 
strong evidence that neutrinos are
massive and lepton flavors are mixed.
Analyses of the atmospheric neutrino deficit in the framework
of either two or three lepton flavors favor 
$\nu_\mu \rightarrow \nu_\tau$ as the dominant oscillation 
mode \cite{Kajita,Fogli} and yield the following
mass-squared difference and mixing factor at the $90\%$ 
confidence level \cite{Kajita}:
\begin{equation}
\Delta m^2_{\rm atm} \sim 10^{-3} ~ {\rm eV^2} \; ,
~~~~~
\sin^2 2\theta_{\rm atm} > 0.9 \; .  
%		(1)
\end{equation}
As for the solar neutrino anomaly,
the hypothesis that solar $\nu_e$ neutrinos
change to $\nu_\mu$ neutrinos during their travel to the earth through
the long-wavelength vacuum oscillation with the parameters
\begin{equation}
\Delta m^2_{\rm sun} \sim 10^{-10} ~ {\rm eV^2} \; ,
~~~~~
\sin^2 2\theta_{\rm sun} \approx 1 \; ,
%		(2)
\end{equation}
can provide a consistent
explanation of all existing solar neutrino data \cite{Barger99}.
Alternatively the large-angle MSW (Mikheyev, Smirnov, and Wolfenstein)
solution, i.e., 
the matter-enhanced $\nu_e\rightarrow
\nu_\mu$ oscillation with the parameters
\begin{equation}
\Delta m^2_{\rm sun} \sim 10^{-5} ~ {\rm eV^2} \; ,
~~~~~~
\sin^2 2\theta_{\rm sun} \sim 1 \; ,
%		(3)
\end{equation}
seems also favored by current data \cite{Bahcall}. To
distinguish between the MSW and vacuum solutions to the
solar neutrino problem is a challenging task of the 
next-round solar neutrino experiments.

Current data indicate that solar and atmospheric neutrino oscillations
are approximately decoupled from each other. Each of them is 
dominated by a single mass scale, i.e.,
\begin{eqnarray}
\Delta m^2_{\rm sun} & = & |\Delta m^2_{21}| \; = \;
\left | m^2_2 - m^2_1 \right | \; , \nonumber \\
\Delta m^2_{\rm atm} & = & |\Delta m^2_{32}| \; = \;
\left | m^2_3 - m^2_2 \right | \; ,
%		(4)
\end{eqnarray}
and $\Delta m^2_{31} \approx \Delta m^2_{32}$ in the scheme
of three lepton flavors. In addition, the $\nu_3$-component in
$\nu_e$ is rather small; i.e., the $V_{e3}$ element of the 
lepton flavor mixing matrix $V$, which links the neutrino mass
eigenstates $(\nu_1, \nu_2, \nu_3)$ to the neutrino flavor
eigenstates $(\nu_e, \nu_\mu, \nu_\tau)$, is suppressed in magnitude. 
Note, however, that the 
hierarchy of $\Delta m^2_{21}$ and $\Delta m^2_{32}$ (or 
$\Delta m^2_{31}$) cannot give clear information about the absolute values
or the relative magnitude of three neutrino masses. For example,
either the strongly hierarchical neutrino mass spectrum 
($m_1 \ll m_2 \ll m_3$) or the nearly degenerate one 
($m_1 \approx m_2 \approx m_3$) is allowed to reproduce the 
``observed'' mass gap between $\Delta m^2_{\rm sun}$ and $\Delta m^2_{\rm atm}$.

In the presence of flavor mixing among three different fermion families,
$CP$ violation is generally expected to appear. This is the case for
quarks, and there is no reason why the same phenomenon does not
manifest itself in the lepton sector \cite{Cabibbo}. 
The strength of $CP$ violation in neutrino oscillations, no
matter whether neutrinos are of the Dirac or Majorana type, 
depends only upon a universal (rephasing-invariant)
parameter $\cal J$, which is defined through
\begin{equation}
{\rm Im} \left (V_{il}V_{jm} V^*_{im}V^*_{jl} \right )
\; =\; {\cal J} \sum_{k,n} \epsilon^{~}_{ijk} \epsilon^{~}_{lmn} 
\; .
%		(5)
\end{equation}
The asymmetry between the probabilities of two $CP$-conjugate 
neutrino transitions, due to the $CPT$
invariance and the unitarity of $V$, are uniquely given as 
\begin{eqnarray}
{\cal A}_{CP} & = & P(\nu_\alpha \rightarrow \nu_\beta) - P(\bar{\nu}_\alpha
\rightarrow \bar{\nu}_\beta) \; \nonumber \\
& = & -16 {\cal J} \sin F_{12} \sin F_{23} \sin F_{31} \; 
%		(6)
\end{eqnarray}
with $(\alpha, \beta) = (e,\mu)$, $(\mu, \tau)$ or $(\tau, e)$,
$F_{ij} = 1.27 \Delta m^2_{ij} L/E$ and
$\Delta m^2_{ij} = m^2_i -m^2_j$, in which $L$ is the distance
between the neutrino source and the detector
(in unit of Km) and $E$ denotes the neutrino beam energy (in unit of
GeV). The $T$-violating asymmetries can be obtained in a
similar way \cite{FX99}:
\begin{eqnarray}
{\cal A}_T & = & P(\nu_\alpha \rightarrow \nu_\beta) - P(\nu_\beta
\rightarrow \nu_\alpha) \; \nonumber \\
& = & -16 {\cal J} \sin F_{12} \sin F_{23} \sin F_{31} \; , 
\nonumber \\ \nonumber \\
{\cal A}'_T & = & P(\bar{\nu}_\alpha \rightarrow \bar{\nu}_\beta) 
- P(\bar{\nu}_\beta \rightarrow \bar{\nu}_\alpha) \; \nonumber \\
& = & +16 {\cal J} \sin F_{12} \sin F_{23} \sin F_{31} \; ,
%		(7)
\end{eqnarray}
where $(\alpha, \beta) = (e, \mu)$, $(\mu, \tau)$ or $(\tau, e)$.
These formulas show clearly that $CP$ or $T$ violation is a behavior
of all three lepton families. In addition,
the relationship ${\cal A}'_T = -{\cal A}_T$ indicates that
the two $T$-violating measurables are odd functions of 
time \cite{Bernabeu}.
A necessary condition for obtaining large (observable) $CP$ or $T$
violation is that the magnitude of $\cal J$ should be large enough.
In view of the smallness of $|V_{e3}|$, one may conclude that the
largeness of $|{\cal J}|$ requires the largeness of both $|V_{e2}|$ and
$|V_{\mu 3}|$. Therefore a mixing scheme which can accommodate
the small-angle MSW solution to the solar
neutrino problem (due to the smallness of $|V_{e2}|$) would not
be able to give rise to large $CP$ or $T$ violation. 

In the following I present a phenomenological model for
lepton mass generation and $CP$ violation within the framework
of three lepton species (i.e., the LSND evidence for neutrino
oscillations, which was not confirmed by the KARMEN 
experiment \cite{Jannakos},
is not taken into account). 
The basic idea, first pointed out by Fritzsch and me in 
1996 \cite{FX96} to get nearly bi-maximal lepton flavor mixing, 
is associated with the flavor democracy
of charged leptons and the mass degeneracy of 
active neutrinos. We 
introduce a simple flavor symmetry breaking 
scheme for charged lepton and neutrino mass matrices, so as to
generate two nearly bi-maximal flavor mixing angles 
and to interpret the approximate decoupling of solar and 
atmospheric neutrino oscillations. Large $CP$ or $T$ violation
of order $|{\cal J}| \sim 1\%$ can naturally emerge in this scenario.
Consequences on the upcoming long-baseline neutrino experiments,
as well as the similarity and difference between lepton and
quark mixing phenomena, will also be discussed.

\vspace{0.5cm}

{\Large\bf 2} ~
Let me start with the symmetry limits of the charged lepton 
and neutrino mass matrices. In a
specific basis of flavor space, in which charged leptons have the
exact flavor democracy and neutrino masses are fully degenerate,
the mass matrices can be written as \cite{FX96,FX98}
\begin{eqnarray}
M^{(0)}_l & = & \frac{c^{~}_l}{3} \left (\matrix{
1	& 1	& 1 \cr
1	& 1	& 1 \cr
1	& 1	& 1 \cr} \right ) \; , 
\nonumber \\
M^{(0)}_\nu & = & c_\nu \left (\matrix{
1	& 0	& 0 \cr
0	& 1	& 0 \cr
0	& 0	& 1 \cr} \right ) \; ,
%		(8)
\end{eqnarray}
where $c^{~}_l =m_\tau$ and $c_\nu =m_0$ measure the 
corresponding mass scales.
If the three neutrinos are of the Majorana type,
$M^{(0)}_\nu$ could take a more general form
$M^{(0)}_\nu P_\nu$ with $P_\nu = {\rm Diag} \{ 1,
e^{i\phi_1}, e^{i\phi_2} \}$. As the Majorana phase matrix
$P_\nu$ has no effect on the flavor mixing and
$CP$-violating observables
in neutrino oscillations, it will be neglected in the subsequent discussions.
Clearly $M^{(0)}_\nu$ exhibits an
S(3) symmetry, while $M^{(0)}_l$ an
$S(3)_{\rm L} \times S(3)_{\rm R}$ symmetry.
In these limits $m_e = m_\mu =0$,
$m_1 = m_2 =m_3 =m_0$, and no flavor mixing is present.

A simple real diagonal breaking of the flavor democracy
for $M^{(0)}_l$ and the mass degeneracy for $M^{(0)}_\nu$
may lead to instructive results for flavor mixing 
in neutrino oscillations \cite{FX96,FX98,Tanimoto}.
To accommodate $CP$ violation, however, complex perturbative
terms are required \cite{FX99}. 
Let me proceed with two different symmetry-breaking steps.

(a) Small real perturbations to the (3,3) elements of $M^{(0)}_l$
and $M^{(0)}_\nu$ are respectively introduced \cite{FH94}:
\begin{eqnarray}
\Delta M^{(1)}_l & = & \frac{c^{~}_l}{3} \left ( \matrix{
0	& 0	& 0 \cr
0	& 0	& 0 \cr
0	& 0	& \varepsilon^{~}_l \cr } \right ) \; , 
\nonumber \\
\Delta M^{(1)}_\nu & = & c_\nu \left ( \matrix{
0	& 0	& 0 \cr
0	& 0	& 0 \cr
0	& 0	& \varepsilon_\nu \cr } \right ) \; .
%		(9)
\end{eqnarray}
In this case the charged lepton mass matrix $M^{(1)}_l =
M^{(0)}_l + \Delta M^{(1)}_l$ remains symmetric under an
$S(2)_{\rm L}\times S(2)_{\rm R}$ transformation, 
and the neutrino mass matrix
$M^{(1)}_\nu = M^{(0)}_\nu + \Delta M^{(1)}_\nu$ has
an $S(2)$ symmetry. 
The muon becomes massive ($m_\mu \approx 2|\varepsilon^{~}_l|
m_\tau /9$), and the mass eigenvalue $m_3$ is no more degenerate
with $m_1$ and $m_2$ (i.e., $|m_3 - m_0| = m_0 |\varepsilon_\nu|$). 
After the diagonalization of 
$M^{(1)}_l$ and $M^{(1)}_\nu$, one finds that the 2nd and 3rd
lepton families have a definite flavor mixing angle
$\theta$. We obtain $\tan\theta \approx -\sqrt{2} ~$ if the
small correction of $O(m_\mu/m_\tau)$ is neglected.
Then neutrino oscillations at the atmospheric scale may arise
from $\nu_\mu \rightarrow \nu_\tau$ transitions with 
$\Delta m^2_{32} = \Delta m^2_{31}
\approx 2m_0 |\varepsilon_\nu|$. The corresponding
mixing factor $\sin^2 2\theta \approx 8/9$ is in good agreement 
with current data.

(b) Small imaginary perturbations, 
which have the identical magnitude but
the opposite signs, 
are introduced to the (2,2) and (1,1) elements of
$M^{(1)}_l$. For $M^{(1)}_\nu$ the corresponding real
perturbations are taken into account \cite{FX99}:
\begin{eqnarray}
\Delta M^{(2)}_l & = & \frac{c^{~}_l}{3} \left ( \matrix{
-i\delta_l	& 0	& 0 \cr
0	& i\delta_l	& 0 \cr
0	& 0	& 0 \cr } \right ) \; , 
\nonumber \\
\Delta M^{(2)}_\nu & = & c_\nu \left ( \matrix{
-\delta_\nu	& ~ 0 	&  0 \cr
0	& ~ \delta_\nu 	&  0 \cr
0	& ~ 0 	&  0 \cr } \right ) \; .
%		(10)
\end{eqnarray}
We obtaine $m_e \approx |\delta_l|^2 m^2_\tau /(27 m_\mu)$ 
and $m_1 = m_0 (1-\delta_\nu)$,
$m_2 = m_0 (1+\delta_\nu)$. The diagonalization of 
$M^{(2)}_l = M^{(1)}_l + \Delta M^{(2)}_l$ and 
$M^{(2)}_\nu = M^{(1)}_\nu + \Delta M^{(2)}_\nu$ 
leads to a full $3\times 3$
flavor mixing matrix, which links neutrino mass eigenstates 
$(\nu_1, \nu_2, \nu_3)$ to neutrino flavor eigenstates
$(\nu_e, \nu_\mu, \nu_\tau)$ in the following manner:
\begin{equation}
V \; = \; \left ( \matrix{
\frac{1}{\sqrt{2}}	& \frac{-1}{\sqrt{2}}	& 0 \cr
\frac{1}{\sqrt{6}}	& \frac{1}{\sqrt{6}}	& \frac{-2}{\sqrt{6}} \cr
\frac{1}{\sqrt{3}}	& \frac{1}{\sqrt{3}}	& \frac{1}{\sqrt{3}} \cr}
\right ) ~ + ~ \Delta V \; 
%		(11)
\end{equation}
with
\begin{equation}
\Delta V \; = \; i ~ \xi^{~}_V ~ \sqrt{\frac{m_e}{m_\mu}} 
\; + \; \zeta^{~}_V ~ \frac{m_\mu}{m_\tau} \; ,
%		(12)
\end{equation}
where
\begin{eqnarray}
\xi^{~}_V & = &  \left ( \matrix{
\frac{1}{\sqrt{6}}	& ~ \frac{1}{\sqrt{6}} ~	& \frac{-2}{\sqrt{6}} \cr
\frac{1}{\sqrt{2}}	& ~ \frac{-1}{\sqrt{2}} ~	& 0 \cr
0	& ~ 0 ~	& 0 \cr} \right ) \; ,
\nonumber \\
\zeta^{~}_V & = & \left ( \matrix{
0	& 0	& 0 \cr
\frac{1}{\sqrt{6}}	& \frac{1}{\sqrt{6}}	& \frac{1}{\sqrt{6}} \cr
\frac{-1}{\sqrt{12}}	& \frac{-1}{\sqrt{12}}	& \frac{1}{\sqrt{3}} \cr}
\right ) \; .
%		(13)
\end{eqnarray}
Some consequences of this mixing scenario can be drawn as follows:

(1) The mixing pattern in Eq. (11), after neglecting small
corrections from the charged lepton masses, is quite similar to that
of the pseudoscalar mesons $\pi^0$, $\eta$ and $\eta'$ in 
QCD \cite{F98}.
One may speculate whether such an analogy could 
be taken as a hint towards an underlying symmetry and its 
breaking, which are responsible for
lepton mass generation and flavor mixing \cite{Wu,Scott}.

(2) The $V_{e3}$ element, of magnitude
\begin{equation}
|V_{e3}| \; =\; \frac{2}{\sqrt{6}} \sqrt{\frac{m_e}{m_\mu}} \; ,
%		(14)
\end{equation} 
is naturally suppressed in
the symmetry breaking scheme outlined above. A similar 
feature appears in the $3\times 3$ quark flavor mixing
matrix, i.e., $|V_{ub}|$ is the smallest among the
nine quark mixing elements \cite{Xing96}. 
Indeed the smallness of $V_{e3}$
provides a necessary condition for the decoupling of
solar and atmospheric neutrino oscillations, even though neutrino
masses are nearly degenerate. The effect of small but nonvanishing
$V_{e3}$ can manifest itself in the long-baseline $\nu_\mu
\rightarrow \nu_e$ and $\nu_e \rightarrow \nu_\tau$ transitions,
as shown in Ref. \cite{FX98}.

(3) The flavor mixing between the 1st and 2nd lepton families
and that between the 2nd and 3rd lepton families are nearly
maximal \cite{FX96}. This property, together with the natural smallness
of $V_{e3}$, allows a satisfactory interpretation of the 
observed large mixing 
in atmospheric and solar neutrino oscillations. We obtain 
\begin{eqnarray}
\sin^2 2\theta_{\rm sun} & = & 1 - \frac{4}{3} \frac{m_e}{m_\mu} 
\; , \nonumber \\
\sin^2 2\theta_{\rm atm} & = & \frac{8}{9} +
\frac{8}{9} \frac{m_\mu}{m_\tau}  \; ,
%		(15)
\end{eqnarray}
to a quite high degree of accuracy. Explicitly
$\sin^2 2\theta_{\rm sun} \approx 0.99$ and $\sin^2 2\theta_{\rm atm}
\approx 0.94$, favored by current data \cite{Kajita}. It is obvious that the model
is fully consistent with the vacuum oscillation solution to
the solar neutrino problem \cite{Barger99} and might also be
able to incorporate the large-angle MSW 
solution \cite{Liu}.

(4) The nearly degenerate neutrino masses and nearly bi-maximal
mixing angles in the present scenario make it possible 
to accommodate the hot dark matter of the 
universe in no conflict with the constraint from the 
neutrinoless double beta decay (denoted by $(\beta\beta)_{0\nu}$).
The former requirement can easily be fulfilled, if one takes
$m_i\sim 1 - 2$ eV (for $i=1,2,3$). The
effective mass term of the $(\beta\beta)_{0\nu}$ decay, in the presence
of $CP$ violation, is written as
\begin{equation}
\langle m \rangle \; =\; \sum_{i=1}^3 \left (m_i U^2_{ei} \right ) \; ,
%		(16)
\end{equation}
where $U = V P_\nu$, and $P_\nu = {\rm Diag} \{ 1, e^{i\phi_1}, e^{i\phi_2} \}$ 
is a diagonal phase matrix of the Majorana nature. Taking
$\phi_1 = \phi_2 =\pi/2$ for example, 
we arrive at 
\begin{equation}
| \langle m \rangle | \; =\; \frac{2}{\sqrt{3}} \sqrt{\frac{m_e}{m_\mu}}
~ m_i \; ,
%		(17)
\end{equation}
i.e., $|\langle m \rangle | \approx 0.08 m_i \leq 0.2$ eV, the latest
experimental bound of the $(\beta\beta)_{0\nu}$ decay \cite{Beta}.
 
(5) The strength of $CP$ violation in this scheme is given by \cite{FX99}
\begin{equation}
{\cal J} \; \approx \; \frac{1}{3\sqrt{3}} \sqrt{\frac{m_e}{m_\mu}}
\left ( 1 + \frac{1}{2} \frac{m_\mu}{m_\tau} \right ) \; .
%		(18)
\end{equation}
Explicitly we have ${\cal J}
\approx 0.014$. The large magnitude of
$\cal J$ for lepton mixing is remarkable, as the same quantity
for quark mixing is only of order $10^{-5}$ \cite{FX95}.
In view of the approximate decoupling of
solar and atmospheric neutrino oscillations,
the $CP$- and $T$-violating asymmetries presented in Eqs. (6) and (7) 
can be simplified to the following form in a long-baseline
neutrino experiment:
\begin{equation}
{\cal A}_{CP} \; = \; {\cal A}_T \; \approx \;
16 {\cal J} \sin F_{12} \sin^2 F_{23} \; .
%		(19)
\end{equation}
We see that the maximal magnitude of
${\cal A}_{CP}$ or ${\cal A}_T$ is able to reach
$16 {\cal J} \approx 0.2$,
significant enough to be measured from the asymmetry
between $P(\nu_\mu \rightarrow \nu_e)$ and 
$P(\bar{\nu}_\mu \rightarrow \bar{\nu}_e)$ 
or that between $P(\nu_\mu \rightarrow \nu_e)$
and $P(\nu_e \rightarrow \nu_\mu)$
in the long-baseline neutrino experiments with the condition
$E/L \sim |\Delta m^2_{12}|$.

Of course the afore-mentioned requirement for the length of the baseline
singles out the large-angle MSW solution, whose oscillation parameters
are given in Eq. (3), among three possible solutions to the solar
neutrino problem. 
If upcoming neutrino oscillation data turn out to rule out 
the consistency between our model
and the large-angle MSW scenario, then it would be quite difficult
to test the model itself from its prediction for
large $CP$ and $T$ asymmetries in the realistic long-baseline
experiments. 

It is at this point worth mentioning that the diagonal non-hermitian 
perturbation to $M^{(0)}_l$ is not the only way to generate
large $CP$ violation in our model. Instead one may consider
the off-diagonal non-hermitian perturbations 
\begin{eqnarray}
\Delta \tilde{M}^{(2)}_l & = & \frac{c^{~}_l}{3} \left ( \matrix{
~ 0 ~	& ~ 0	& i\delta_l \cr
~ 0 ~	& ~ 0	& -i\delta_l \cr
~ i\delta_l ~	& -i\delta_l	& 0 \cr } \right ) \; ,
\nonumber \\
\Delta \hat{M}^{(2)}_l & = & \frac{c^{~}_l}{3} \left ( \matrix{
-i\delta_l	& 0	& i\delta_l \cr
0	& i\delta	& -i\delta_l \cr
i\delta_l	& -i\delta_l	& 0 \cr } \right ) \; ;
%		(20)
\end{eqnarray}
or the off-diagonal hermitian perturbations
\begin{eqnarray}
\Delta {\bf M}^{(2)}_l & = & \frac{c^{~}_l}{3} \left ( \matrix{
~ 0 ~	& -i\delta_l ~	& ~ 0 ~~ \cr
~ i\delta_l ~	& ~ 0 ~	& ~ 0 ~~ \cr
~ 0 ~	& ~ 0 ~	& ~ 0 ~~ \cr } \right ) \; , 
\nonumber \\
\Delta \tilde{\bf M}^{(2)}_l & = & \frac{c^{~}_l}{3} \left ( \matrix{
~ 0	& ~ 0 ~	& i\delta_l \cr
~ 0	& ~ 0 ~	& -i\delta_l \cr
- i\delta_l	&  ~ i\delta_l ~	& 0 \cr } \right ) \; ,
\nonumber \\
\Delta \hat{\bf M}^{(2)}_l & = & \frac{c^{~}_l}{3} \left ( \matrix{
0	& -i\delta_l	& i\delta_l \cr
i\delta	& 0	& -i\delta_l \cr
-i\delta_l	& i\delta_l	& 0 \cr } \right ) \; ,
%		(21)
\end{eqnarray}
for the same purpose. Note that all the six 
perturbative mass matrices 
$\Delta M^{(2)}_l$,
$\Delta \tilde{M}^{(2)}_l$,
$\Delta \hat{M}^{(2)}_l$ and
$\Delta {\bf M}^{(2)}_l$,
$\Delta \tilde{\bf M}^{(2)}_l$,
$\Delta \hat{\bf M}^{(2)}_l$
have a common feature: the (1,1) elements of their 
counterparts in the hierarchical basis all vanish \cite{FX99}.
This feature assures that the $CP$-violating effects, resulted from
the above complex perturbations, 
are approximately independent 
of other details of the flavor symmetry breaking and have
the identical strength to a high degree of accuracy.
Indeed it is easy to check that the relevant
charged lepton mass matrices, 
together with the neutrino mass matrix
$M^{(2)}_\nu = M^{(0)}_\nu + \Delta M^{(1)}_\nu +
\Delta M^{(2)}_\nu$, lead to the same flavor mixing pattern $V$
as given in Eq. (11) \cite{FX99}. Hence it is in practice
difficult to distinguish one scenario from
another. In our point of view, the simplicity of $M^{(2)}_l$
and its parallelism with $M^{(2)}_\nu$ might make it 
technically more natural to be derived from a yet unknown
fundamental theory of lepton mixing and $CP$ violation.

\vspace{0.5cm}

{\Large\bf 3} ~
In the scheme of three lepton species
I have introduced a simple phenomenological
model for lepton mass 
generation, flavor mixing and $CP$ violation. The 
model starts from the flavor democracy of charged leptons
and the mass degeneracy of neutrinos. After the 
symmetry limits of both mass matrices are explicitly broken, 
we find that this scenario can naturally
give rise to large flavor mixing and large $CP$ or $T$ violation. The
condition for the approximate decoupling of atmospheric and
solar neutrino oscillations (i.e., $|V_{e3}| \ll 1$) is
also fulfilled. 

The lepton mixing pattern, which arises from $m_e \ll m_\mu \ll m_\tau$
and $m_1 \approx m_2 \approx m_3$, seems to be ``anomalously'' 
different from the quark mixing pattern obtained from the fact
$m_u \ll m_c \ll m_t$ and $m_d \ll m_s \ll m_b$. However, their
similarities do exist. To see this point clearly, let me parametrize
the quark (Q) and lepton (L) flavor mixing matrices in an
instructive way \cite{FX97}:
\begin{eqnarray}
V_{\rm Q} & = & R_{12}(\theta_{\rm u},0) R_{23}(\theta_{\rm Q}, -\phi_{\rm Q}) 
R^{\rm T}_{12}(\theta_{\rm d},0) \; , \nonumber \\
V_{\rm L} & = & R_{12}(\theta_{l}, 0) R_{23}(\theta_{\rm L}, -\phi_{\rm L})
R^{\rm T}_{12}(\theta_{\nu},0) \; ,
%		(22)
\end{eqnarray}
where the complex rotation matrices $R_{12}$ and $R_{23}$ are defined as
\begin{eqnarray}
R_{12}(\theta, \phi) & = &
\left ( \matrix{
c	& s	& 0 \cr
-s	& c	& 0 \cr
0	& 0	& e^{i\phi} \cr} \right ) \; ,
\nonumber \\
R_{23}(\theta, \phi) & = & 
\left ( \matrix{
e^{i\phi}	& 0	& 0 \cr
0	& c	& s \cr
0	& -s	& c \cr} \right ) \; 
%		(23)
\end{eqnarray}
with $c \equiv \cos\theta$ and $s \equiv \sin\theta$.
Note that the rotation sequence of $V_{\rm Q}$
or $V_{\rm L}$ is essentially the original Euler sequence
with an additional $CP$-violating phase. 
The rotation angle $\theta_l$ (or $\theta_\nu$) mainly
describes the mixing between $e$ and $\mu$ leptons (or between
$\nu_e$ and $\nu_\mu$ neutrinos), and the rotation angle $\theta_{\rm u}$
(or $\theta_{\rm d}$) primarily describes the mixing between
$u$ and $c$ quarks (or between $d$ and $s$ quarks). The rotation
angle $\theta_{\rm Q}$ (or $\theta_{\rm L}$) 
is a combined effect arising from the mixing between
the 2nd and 3rd families for quarks (or leptons). The
phase parameters $\phi_{\rm Q}$ and $\phi_{\rm L}$
signal $CP$ violation in flavor mixing
(for neutrinos of the Majorana type, two additional $CP$-violating
phases may enter but they are irrelevant for neutrino oscillations).
Comparing Eqs. (11)--(13) with (22) we immediately arrive at
\begin{equation}
\tan\theta_l \; =\; \sqrt{\frac{m_e}{m_\mu}} \; ,
~~~~~~~ \tan\theta_\nu \; =\; \sqrt{\frac{m_1}{m_2}} \; . \;\;\;\;\;
%		(24)
\end{equation}
In contrast, a variety of quark mass matrices have predicted \cite{FX95,F78}
\begin{equation}
\tan\theta_{\rm u} \; =\; \sqrt{\frac{m_u}{m_c}} \; ,
~~~~~~~ \tan\theta_{\rm d} \; =\; \sqrt{\frac{m_d}{m_s}} \; .
%		(25)
\end{equation}
As one can see, the large mixing angle $\theta_\nu$ (i.e.,
$\tan\theta_\nu \approx 1$ for almost degenerate $m_1$ and $m_2$)
is attributed
to the near degeneracy of neutrino masses in our flavor symmetry
breaking scheme. Explicitly three mixing angles of leptons 
take the values \cite{FX99}
\begin{equation}
\theta_l \approx 4^{\circ} \; , ~~~
\theta_\nu \approx 45^{\circ} \; , ~~~
\theta_{\rm L} \approx 52^{\circ} \; ; 
%		(26)
\end{equation}
and those of quarks take the values \cite{Stocchi}
\begin{equation}
\theta_{\rm u} \approx 5^{\circ} \; , ~~~
\theta_{\rm d} \approx 11^{\circ} \; , ~~~
\theta_{\rm Q} \approx 2^{\circ} \; .
%		(27)
\end{equation}
Furthermore, the $CP$-violating phases 
$\phi_{\rm L}$ and $\phi_{\rm Q}$ are both close to a special
value:
\begin{equation}
\phi_{\rm Q} \; \approx \; \phi_{\rm L} \; \approx \; 90^{\circ} 
\; .
%		(28)
\end{equation}
Let me emphasize that the possibility $\phi_{\rm Q} \approx 90^{\circ}$
is favored by a variety of realistic 
models of quark mass matrices \cite{FX95}. The result 
$\phi_{\rm L} \approx 90^{\circ}$ is a distinctive feature of our 
lepton mixing scenario, but to verify it in a model-independent way
would be extremely difficult, if not impossible, in the future
neutrino experiments. Indeed the questions, about how large
the feasibility is and how much the cost will be to
measure $CP$ or $T$ violation in neutrino 
oscillations \cite{Cabibbo,FX99,Bernabeu,CP}, remain open.

We are expecting that more data from the Super-Kamiokande and other neutrino
experiments could provide stringent tests of the
model discussed here.

\vspace{0.5cm}
{\it Acknowledgments} ~ This talk is based on the works in
collaboration with H. Fritzsch. I am indebted to him for his
constant encouragement.
I would like to thank F.L. Navarria
and the organizing committee of {\it San Miniato 1999}
for partial financial support.
I am grateful to Q.Y. Liu, W.G. Scott and Y.L. Wu for intensive
and interesting discussions during the workshop.

\end{document}